# Analytical Study for Seeking Relation Between Customer Relationship Management and Enterprise Resource Planning

Asif Perwej
asifperwej@gmail.com
Singhania University, Rajsthan, India

**Abstract**
Enterprise Resource Planning (ERP) is a integration of various resources of any organization. It is computer software. All kinds of organization data that is relating to each and every function of the organization are available in ERP. So most of the big business organizations are implementing ERP and some of the medium, small scale companies are also using ERP system. CRM in an organization helps to retain their existing customers as well as capturing new customers for their products. So it makes the organization to produce those goods required by their consumers. This paper focuses mainly on the merging of CRM and ERP through Neural Networks.

## 1. Introduction

ERP consists of all data which are relating to the different functions of the organization. In the Competitive world today Organizations are developing both horizontal as well as vertical wise. This paves way for accumulation of larger amount of data. By following the older method of DBMS is not enough to keep the large volume of data's. So ERP came into existence. It helps to keep the entire growing data under one system and provides all the departmental information as and when required by the concerned department. Further ERP provides valuable suggestions in decision making process. ERP system has three major parts that is client server system, database and application part or end user part (Yen et al., 2001). ERP system is operated on the basis of client server technology. Its applications are deployed in distributed and dispersed system. Some ERP systems provide web interfaces. ERP system is always implemented through many databases. All applications in the ERP system interact with the database, which ensures the data integrity of the enterprise. ERP provides enhancements in different fields of customer relationship management such as, production, sales and distribution. It also helps the managers to take timely decision makings since it is the integration of all functional departments (McCarthy, 2000and James, 2002). It is difficult to implement successfully since it has huge amount of information in relation with all the departments (Wen-Hsiung Wu, Chin-Fu Ho, Hsin-Pin Fu, Tien-Hsiang Chang). ERP handles all kinds of problems related with all the functions of the organization and provides good solutions (Gail Corbitt, Marinos themistocleous, Zahir Irani, 2005).

## 2. CUSTOMER RELATIONSHIP MANAGEMENT

Customer Relationship Management (CRM) comprises sequence of steps to be followed for keeping the customers for longer period of time. It is also useful to maintain a healthy relationship between the organization and the consumers (Ling & Yen, 2001).

According to (Ling and Yen, 2001) and (Ngai, 2005). Swift (2001, p. 12) CRM maintains proper communication between the organization and the consumers which enables the organization to understand the feelings of the consumers. It helps the companies to get new customers, retaining the existing customers and increases the profits of the business. Since CRM helps to create intimate relationship with the consumers, the company knows what exactly the consumers want from the company in terms of product. (Peppers, Rogers & Dorf, 1999). The primary goal of CRM is to retain the old customers as long as possiblein the industry (Peppard, 2000). Nowadays, in a modern society most of the business transactions are to be done through online. In return the organizations get their customers and retain them is called as web marketing(Schafer, Konstan, & Riedl, 2001).

## 3. NEURAL NETWORKS

Artificial Neural Networks (ANNs) are distributedand parallel information systems. It simulates the human brain to process information. It is a collective system that comprises of neurons and nodes. The nodesare connected through direct links.

6

One or more inputvalues are taken and they are combined as a singlevalue, processed and produced as an output value. ANNis used in case of examining the complex relationship between the input and output variables (Nelson & Illingworth, 1994). ANN is used mostly in commercial fields like market segmentation, sales forecasting, direct marketing, new product development, and target marketing (Bishop, 1995; Callan, 1999; Curry & Moutinho, 1993; Fausett, 1994; Hassoum, 1995; Hu, Shanker, & Hung, 1999; Kim, Street, Russell,135 *International Journal of Computer Communication and Information System ( IJCCIS)– Vol2. No1. ISSN: 0976–1349 July – Dec 2010* &Menczer, 2005; Zahavi & Levin, 1997; Zhang, Hu,Patuwo, & Indro, 1999).

Nowadays it is applied in the area of customer relationship management (Audrain, 2002; Hackl & Westlund, 2000; Willson & Wragg, 2001). ANN is used to find out consumer behavior and is more suitable in this area than that of any other statistical methods (Baesens et al., 2002; Viaene et al., 2002). ANN is very much useful in where the data is abnormal and having no linear relationships between them. It gives more accurate results than multiple regression Models (Bishop 1994).

## 4. Studying relation between ERP and CRM

The integration of ERP and CRM produces the best results to the organizations. Any organization usually receives the customer complaints or comments first forgetting feedback from them. This department is known as Customer relationship management department. Then the complaints are to be sent to the respective department. They are handled by ERP department and necessary actions are to be taken here. All the complaints received and the respective actions taken are to be stored in the database. This is known as knowledge discovery. Data mining techniques are used here to dig valuable information for the management as a whole as well as for the CRM department. The Neural networks technique is one of the best choices among the data mining techniques for integrating ERP and CRM.

## 5. CONCLUSION

This paper focuses that the data mining technique neural networks can be used for further research in ERP-CRM integration. This can be used in any kind of industries which has CRM and ERP departments. It helps to take valuable and optimal decisions on the customer side as well as for the entire management system.